\documentclass[aps,prx,twocolumn,superscriptaddress,showpacs,nofootinbib]{revtex4-1}

\usepackage{graphicx,lipsum}
\usepackage{amsmath,color}
\usepackage{pgfplots}

\usepackage[normalem]{ulem}

\def\gappeq{\mathrel{ \rlap{\raise.5ex\hbox{$>$}}
                      {\lower.5ex\hbox{$\sim$}} } }
\def\lappeq{\mathrel{ \rlap{\raise.5ex\hbox{$<$}}
                      {\lower.5ex\hbox{$\sim$}} } }
\usepackage{bm}
\usepackage{color}
\usepackage{epstopdf}

\newcommand\edd{\varepsilon_\text{dd}}
\newcommand{\del}[1]{\textcolor{red}{}}

\newcommand\di{\text{d}}
\graphicspath{{figs/}}

\begin{document}

\title{Anomalous oscillations of dark solitons in trapped dipolar condensates}

\author{T. Bland} 
\affiliation{Joint Quantum Centre Durham--Newcastle, School of Mathematics and Statistics, Newcastle University, Newcastle upon Tyne, NE1 7RU, United Kingdom}
\author{K. Paw\l{}owski}
\affiliation{Center for Theoretical Physics, Polish Academy of Sciences, Al. Lotnik\'ow 32/46, 02-668 Warsaw, Poland}
\author{M. J. Edmonds}
\affiliation{Joint Quantum Centre Durham--Newcastle, School of Mathematics and Statistics, Newcastle University, Newcastle upon Tyne, NE1 7RU, United Kingdom}
\author{K. Rz\c{a}\.{z}ewski}
\affiliation{Center for Theoretical Physics, Polish Academy of Sciences, Al. Lotnik\'ow 32/46, 02-668 Warsaw, Poland}
\author{N. G. Parker}
\affiliation{Joint Quantum Centre Durham--Newcastle, School of Mathematics and Statistics, Newcastle University, Newcastle upon Tyne, NE1 7RU, United Kingdom}


\begin{abstract}
Thanks to their immense purity and controllability, dipolar Bose-Einstein condensates are an examplar for studying fundamental non-local nonlinear physics.  Here we show that a family of fundamental nonlinear waves - the dark solitons - are supported in trapped quasi-one-dimensional dipolar condensates and within reach of current experiments.  Remarkably, the oscillation frequency of the soliton is strongly dependent on the atomic interactions, in stark contrast to the non-dipolar case.  The failure of a particle analogy, so successful for dark solitons in general, to account for this behaviour implies that these structures are inherently extended and non-particle-like.  These highly-sensitive waves may act as mesoscopic probes of the underyling quantum matter field.
\end{abstract}

\maketitle

Dark solitons are the fundamental nonlinear excitations of one-dimensional media with defocussing nonlinearity, appearing as a travelling localized reductions in the field amplitude.  Since first realized in optical fibres \cite{Emplit,Krokel,Weiner}, they have been observed across plasmas \cite{Shukla,Heidemann}, water \cite{Chabchoub}, magnetic films \cite{Chen} and atomic Bose-Einstein condensates (BECs) \cite{Burger1999,Denschlag2000,Dutton2001,anderson_2001,Jo2007,Engels2007,Chang2008,Becker2008,Weller2008,Stellmer2008,Chang2008,shomroni_2009,hamner_2011,Aycock2016}.  
The latter system provides a commanding playground for exploring soliton physics in which the nonlinearity (viz. atomic interactions) can be precisely controlled in amplitude, time and space \cite{Bloch2008}, and almost arbitrary potentials can be painted \cite{Henderson2009}.   Experiments have studied a host of fundamental properties, including their collisions~\cite{Stellmer2008,Weller2008}, creation ~\cite{Burger1999,Denschlag2000,Engels2007,Jo2007,Chang2008}, interaction with impurities \cite{Aycock2016}, and decay ~\cite{anderson_2001,shomroni_2009}.   Moreover, these ``quantum canaries" are touted as sensitive probes of the mesoscale quantum physics within the quantum degenerate gas \cite{Anglin2008}.

It is remarkable that the dark soliton, a collective excitation, behaves to first order as a classical particle with negative effective mass, acting under the external potential \cite{kivshar_1998,Frantzeskakis2010}.  For example, in harmonically-trapped BECs, the soliton oscillates at a characteristic ratio, $\omega/\sqrt{2}$, of the trap frequency $\omega$ \cite{busch_2000,Huang2002,konotop_2004,brazhnyi_2006,pelinovsky_2005,frantzeskakis_2007,Theocharis2007,pelinovsky_2008,kamchatnov_2009,astrakharchik_2013}, as confirmed experimentally \cite{Weller2008}.  This robust result, insensitive to the microscopic atomic interactions, is a signature of matter-wave dark solitons.  Here we establish the form and dynamics of these fundamental structures in trapped BECs featuring dipole-dipole atomic interactions.  Remarkably, the oscillations become strongly dependent on the strength and polarization of the dipolar interactions. The dynamics cannot be accounted for within the particle analogy, implying that the dark solitons are strictly extended, non-particle-like excitations.  We establish these solutions and their oscillatory behaviour, based on one- and three-dimensional mean-field models, and demonstrate that they are accessible to current experiments.

The last decade has seen a surge of research on dipolar BECs,as realized through the condensation of vapours of $^{52}$Cr \cite{griesmaier_2005,beaufils_2008}, $^{164}$Dy \cite{lu_2011,tang_2015} and $^{168}$Er \cite{aikawa_2012}.  On top of the usual van der Waals (vdW) interatomic interactions, which are isotropic and short-range, the atoms possess significant magnetic dipole moments and experience dipole-dipole (DD) interactions, which are anisotropic and long-range~\cite{lahaye_2009}.  This has opened the door to study the interplay of magnetism with quantum coherence, and local with non-local nonlinearities, at the control of atomic physics.  Rich phenomena have been revealed, including recent observations of the quantum analog of the ferrofluid Rosensweig instability \cite{kadau_2016,ferrier_2016} and self-bound three-dimensional droplets \cite{chomaz_2016,schmitt_2016}.  

%
%

We consider a trapped, weakly-interacting BEC of atoms with mass $m$ and permanent magnetic dipole moment $\mu$, polarized in a common direction, and in the limit of zero temperature.   
The atom-atom interactions can be approximated by the universal pseudo-potential \cite{lahaye_2009}
\begin{eqnarray}
U(\textbf{r}-\textbf{r}')=\frac{4 \pi \hbar^2 a_s}{m}\delta(\textbf{r}-\textbf{r}')+\frac{\mu_0 \mu^2}{4\pi}\frac{1-3\cos^2\Theta}{|\textbf{r}-\textbf{r}'|^3}.
\label{eqn:U}
\end{eqnarray}
The first term describes the vdW interactions, characterised by the {\it s}-wave scattering length $a_s$; this is experimentally tunable through Feshbach resonances under external magnetic or optical fields \cite{chin_2010}. The second term is the DD interaction, where $\mu_0$ is the permeability of free space and $\Theta$ is the angle between the inter-atom vector and the polarization direction.  It is useful to define the dipolar lengthscale $a_{\rm dd}=m \mu_0 \mu^2/12 \pi \hbar^2$.   The magic angle $\Theta_m\approx54^\circ$, for which this term reduces to zero, is the crossover from attractive to repulsive DD interactions.  For $\Theta>\Theta_m$ the dipoles repel while for $\Theta<\Theta_m$ they attract.  The regime of ``anti-dipoles", $\mu^2<0$, is accessible by tilting and rapidly rotating the polarization direction \cite{giovanazzi_2002}; then this angular behaviour becomes reversed. We quantify the interactions through the relative interaction parameter $\edd=a_\text{dd}/a_s$ \cite{lahaye_2009}, where the full range $-\infty<\edd<\infty$ is experimentally accessible.


The trapping potential is assumed to be harmonic and axi-symmetric, $V=m [\omega_z^2 z^2+\omega_\perp^2 r^2]/2$, where $\omega_z$ and $\omega_\perp$ are the axial and radial trap frequencies, respectively.
The polarization is at an angle $\theta$ to the $z$-axis.
The BEC is described by a (complex) mean-field wavefunction $\Psi(\textbf{r},t)$, 
normalized to the number of atoms, $N$, which obeys the dipolar Gross-Pitaevskii equation \cite{lahaye_2009}, 
\begin{eqnarray}
i\hbar\frac{\partial \Psi}{\partial t}&=&\left[-\frac{\hbar^2}{2m}\nabla^2+\frac{m}{2}(\omega_z^2z^2+\omega_\perp^2 r^2)+\frac{4 \pi \hbar^2 a_s}{m}|\Psi|^2 \right. \nonumber
\\
&~& \left.+ \int U_{\rm dd}({\bf r}-{\bf r}')|\Psi(\textbf{r}',t)|^2 ~\di\textbf{r}'\right]\Psi,
\label{eqn:dgpe3d}
\end{eqnarray}
where $U_{\rm dd}$ denotes the DD term in Eq. (\ref{eqn:U}).  In effect, the BEC experiences an effective potential comprising of the static external potential, a local potential proportional to the atomic density arising from vdW interactions, and a non-local potential arising from the DD interactions. 

Since dark solitons are dimensionally unstable in 3D (decaying into vortical structures via the snake instability), we focus on highly-elongated BECs.  First, for simplicity, we work deep in this quasi-1D limit ($\omega_z \ll \omega_\perp$ and $\hbar \omega_\perp > \mu$, where $\mu$ is the BEC chemical potential).  The 3D wavefunction $\Psi$ then approximates the form $\Psi(\textbf{r},t)=\psi_\perp(x,y)\psi(z,t)$ where $\psi_\perp(x,y)=(l_\perp\sqrt{\pi})^{-1}\exp\{-(x^2+y^2)/2l_\perp^2\}$ is the transverse ground harmonic oscillator state with characteristic length $l_\perp=\sqrt{\hbar/m\omega_\perp}$.  Integrating out the transverse mode leads to an effective 1D dipolar GPE \cite{sinha_2007,deuretzbacher_2010}, equivalent to Eq. (\ref{eqn:dgpe3d}) under the replacements ${\bf r} \mapsto z$, $\Psi \mapsto \psi$, $a_s \mapsto a_s/2\pi \l_\perp^2$ and 
\begin{equation}
U_{\text{dd}} \mapsto U_0 \bigg[2u{-}\sqrt{2\pi}(1{+}u^2)e^{u^2/2}\text{erfc}\bigg(\frac{u}{\sqrt{2}}\bigg)+\frac{8}{3}\delta(u)\bigg], \nonumber
\label{eqn:pp1d}
\end{equation}
where $u=|z-z'|/l_\perp$ and $U_0=\mu_0 \mu^2 (1+3\cos2\theta)/{32\pi l_\perp^3}$.

In the absence of dipoles and axial trapping, and for repulsive vdW interactions ($g>0$), the 1D dipolar GPE reduces to the 1D defocussing cubic nonlinear Schr\"odinger equation.  This  is completely integrable, supporting a family of dark soliton solutions \cite{zakharov_1973,kivshar_1998} with characteristic density depresssion and phase slip. 
Axial trapping and/or dipolar interactions break this integrability but continue to support dark solitons (defined broadly) which may be found numerically \cite{busch_2000,Frantzeskakis2010,pawlowski_2015,bland_2015,edmonds_2016}.  Bright \cite{cuevas_2009,adhikari_2014,baizakov_2015,umarov_2016,edmonds_2016b} and bright-dark \cite{adhikari_2014b} solitons have been predicted in dipolar BECs, although these are quite distinct from dark solitons.

We can specify a criterion for dark soliton to exist in the dipolar BEC.  Within the local density approximation (LDA), the interaction terms in the 1D dipolar GPE reduce to $\hbar \omega_\perp[2 a_s - a_{\rm dd}  (1+3 \cos 2 \theta)/2] n(z)$, where $n(z)=|\psi|^2$ is the axial density profile. Enforcing these net interactions to be repulsive (positive) leads to the rudimentary criterion to support dark solitons,
\begin{equation}
a_{\rm eff}=a_s\left[1+\frac{\edd}{2}(1-3 \cos^2 \theta) \right]>0,
\label{eqn:criterion}
\end{equation} 
where $a_{\rm eff}$ is an effective {\it s}-wave scattering.

\begin{figure}[t]
\centering
\includegraphics[scale=1]{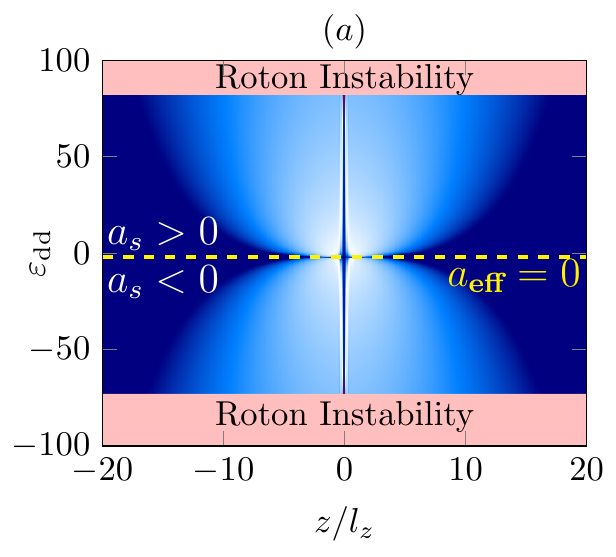}
\includegraphics[scale=1]{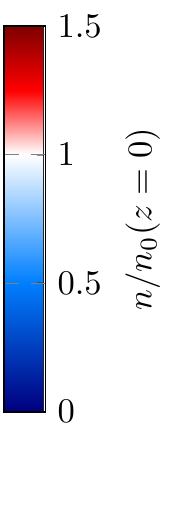}
\includegraphics[scale=1]{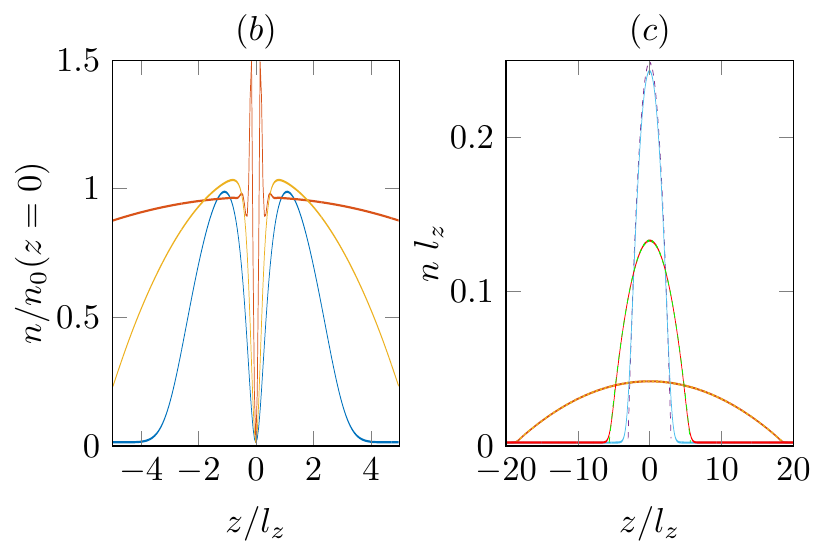}
\caption{(Colour online) (a) Density profile $n(z)$ of the quasi-1D dipolar BEC (polarization perpendicular to the axis) featuring a central black soliton, as a function of $\edd$.  The vdW interactions satisfy $|\beta|=61$ and the trap frequency ratio $\omega_z/\omega_\perp=0.0025$. 
Only the regimes satisfying Eq. (\ref{eqn:criterion}) are shown, with the line $a_{\rm eff}=0$ indicated (yellow dashed line).  The color scale is normalised to the peak density of the soliton-free BEC, $n_0(z=0)$.  The roton-unstable regions extend to $\edd = \pm \infty$.  (b) Example density profiles, for $\edd=-1.7$ (blue lines), $\edd=-74$ (red lines) and $\edd=0$ (yellow lines).  (c) Soliton-free density profile (solid lines), with the TF prediction of Eq. (\ref{eqn:diptf}) overlaid (dotted), for the same $\edd$ values as in (b).}
\label{fig:spec}
\end{figure}

We illustrate the dark soliton solutions using the case of dipoles polarized perpendicular to the $z$-axis ($\theta=\pi/2$).  The criteria (\ref{eqn:criterion}) then reduces to $a_s(1+\edd/2)>0$, or, in terms of $\edd$,  $\edd>-2$ for $a_s>0$ and $\edd<-2$ for $a_s<0$.  We only consider the solutions in these ranges; elsewhere the BEC has net attractive interactions and does not support dark solitons. Figure~\ref{fig:spec}$(a)$ maps the density $n(z)$ of the quasi-1D BEC featuring a central black soliton, as a function of $\edd$.  The vdW interactions, characterised by the dimensionless parameter $\beta=a_sNl_z/l_\perp^2$, are fixed in amplitude throughout to the nominal value $|\beta|=61$.   This black soliton state corresponds to the first excited state of the BEC \cite{Frantzeskakis2010}, and is obtained by numerical integration of the 1D dipolar GPE in imaginary time (using a Crank-Nicolson scheme) subject to a $\pi$-phase step at the origin.

The background BEC widens as $\edd$ is varied away from the line $a_{\rm eff}=0$, caused by magnetostriction in the playoff between the vdW and DD interactions.  This can be accounted for within the Thomas-Fermi approximation, valid for strong repulsive interactions and based on neglecting density gradients.  Generalizing previous derivations of the Thomas-Fermi profile of the quasi-1D trapped BEC \cite{menotti_2002,parker_2008} to include dipoles aligned at an arbitrary angle $\theta$ leads to the Thomas-Fermi density  \cite{Cai2010},
\begin{eqnarray}
n_{\rm TF}(z)=\frac{(l_\perp R_z/2l_z^2)^2}{a_{\rm eff}}\left[1-\frac{z^2}{R_z^2}\right],
\label{eqn:diptf}
\end{eqnarray}
for $z \leq R_z$, and $n_{\rm TF}=0$ elsewhere, where $R_z=[3a_{\rm eff}Nl_z^4/l_\perp^2]^{1/3}$ defines the Thomas-Fermi half-width of the BEC.  The angular dependence is intuitive: for axially-polarized dipoles (perpendicularly-polarized), $R_z$ is smaller (larger) than the non-dipolar case, consistent with a head-to-tail (side-by-side) alignment shrinking (enlarging) the axial extent of the cloud.   The TF prediction typically agrees very well with the true profiles [see Fig.~\ref{fig:spec}$(c)$], with significant deviations only when the net local interactions become small ($a_{\rm eff} \rightarrow 0$).  

The background BEC suffers the roton instability (RI).  A trapped dipolar BEC can develop a roton (finite-momentum) minimum in its excitation spectrum which, for certain parameters, can touch zero energy, triggering an instability at finite momentum \cite{Bohn2009}.  Our quasi-1D BEC has three RI regimes. The first also arises in the uniform system, as mapped out elsewhere \cite{bland_2015,edmonds_2016}; e.g., in Fig. \ref{fig:spec}(a) this occurs for $\edd \gappeq -2$ with $a_s<0$.  Two further RI regimes arise for large $|\edd |$ (red bands in Fig. \ref{fig:spec}(a)).  These RI bands encroach towards $\edd=0$ as the system becomes more 3D (increasing trap ratio $\omega_z/\omega_\perp$).

The black soliton appears as a density notch at the origin, set upon the background BEC.  For $\edd=0$ and with $a_s>0$ the numerical solution (yellow line in Fig. \ref{fig:spec}(b)) closely approximates the product of the exact black soliton solution in a uniform system \cite{zakharov_1973,kivshar_1998} and the background density $n_{\rm b}(z)$, i.e. $n(z)=n_{\rm b}(z)\tanh^2 (z/\xi)$, where $\xi=1/\sqrt{4 \pi  n_0 a_s}$ is the healing length at the BEC centre.  For $\edd \neq 0$, and away from $a_{\rm eff}=0$ and the RIs, this approximate form holds, with $a_s$ replaced by $a_{\rm eff}$.  However, close to a RI the dark soliton develops distinctive peripheral density ripples (e.g. red line in  Fig. \ref{fig:spec}(b)), due to the mixing of the roton mode into this state, and as reported in the uniform system \cite{pawlowski_2015,bland_2015,edmonds_2016}.  Meanwhile, as $a_{\rm eff}=0$ is approached, the soliton broadens while the background BEC shrinks.   While we have focussed on $\theta=\pi/2$, the behaviour is qualitatively similar for all $\theta > \theta_{\rm m}$ (where $\theta_{\rm m}$ is the magic angle relative to the $z$-axis), albeit with shifts in $a_{\rm eff}$ (according to Eq. (\ref{eqn:criterion})) and the onsets of the RI.  Meanwhile, for $\theta<\theta_{\rm m}$ the dependence on $\edd$ is effectively flipped \cite{bland_2015,edmonds_2016}.

\begin{figure}[!h]
	\centering
	\includegraphics[scale=1]{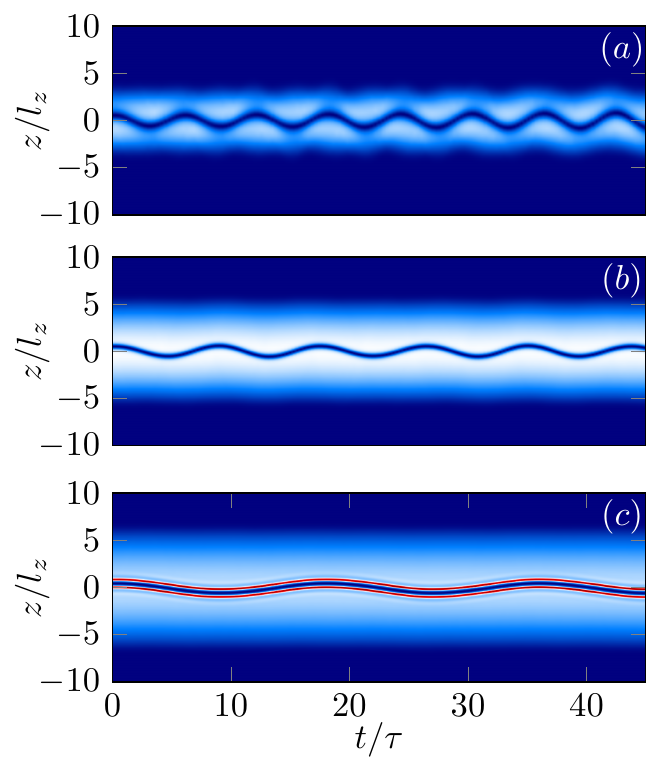}
	\includegraphics[scale=1]{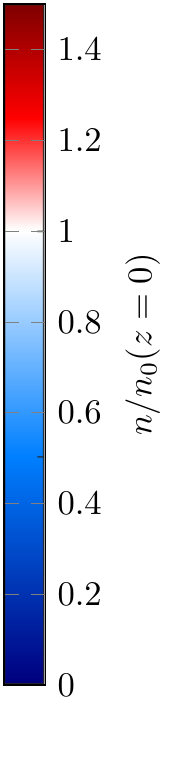}
	\caption{(Color online) Density dynamics of a dark soliton in the quasi-1D dipolar condensate for (a) $\edd=-1.7$, (b) $\edd=0$, and (c) $\edd=-5.5$.  These values correspond to close to $a_{\rm eff}=0$, the non-dipolar case, and close to the roton instability, respectively.  Remaining parameters as in Fig. \ref{fig:spec}.}
	\label{fig:osceg05}
\end{figure}

\begin{figure}[t]
	\centering
	\includegraphics[scale=0.9]{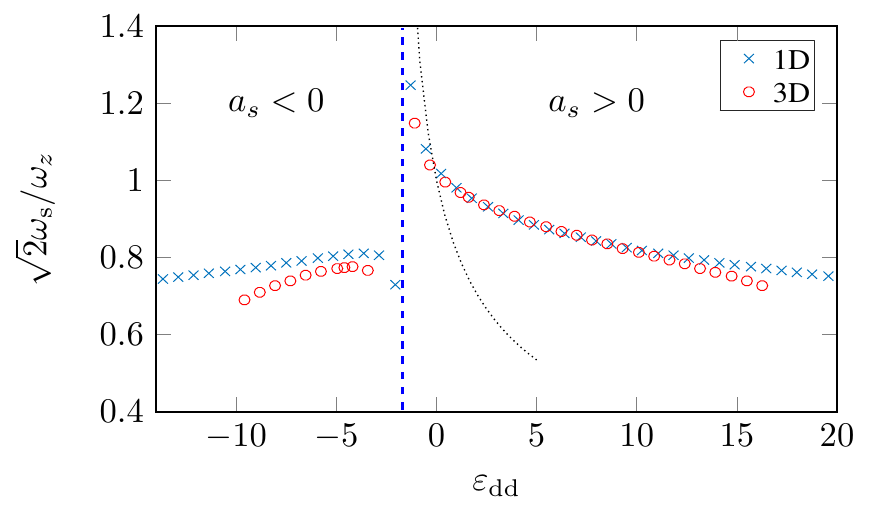}
	\caption{(Color online) Oscillation frequency of the dark soliton (starting as an off-centre black soliton) based on the 1D dipolar GPE (blue crosses), 3D dipolar GPE (red circles), and $\omega_{z,{\rm eff}}/\sqrt{2}$ (black dotted line). The system parameters are as Fig. \ref{fig:spec}; the 3D system also assumes $^{164}$Dy atoms, $\omega_{\perp} = 2 \pi \times 16$~kHz and $|a_s| = 50a_0$, where $a_0$ is the Bohr radius.  The 3D system is stable only in the range of markers.   } 
	\label{fig:freqsig05}
\end{figure}

Next we study the oscillation dynamics of the dark soliton, from the initial condition of an off-centre black soliton (at $z=0.5 l_z$, although our findings are insensitive to the initial offset).  Identical results are obtained by using the product of the background BEC and a travelling dark soliton solution from the uniform system \cite{bland_2015,edmonds_2016}.  Figure \ref{fig:osceg05} shows three example cases with differing $\edd$ (close to $a_{\rm eff}=0$, the non-dipolar case $\edd=0$ and close to a RI).  Throughout, the soliton oscillates sinusoidally and stably through the BEC, with preserved form and oscillation amplitude.  It is clear, however, that the oscillation period changes with $\edd$ (even for condensates with similar sizes and curvatures, c.f. Figs. \ref{fig:osceg05}(b) and (c)).  To explore this further, Fig. \ref{fig:freqsig05} plots the oscillation frequency $\omega_{\rm s}$ of the soliton coordinate (defined as the point of minimum density) based on the 1D dipolar GPE (blue crosses).  For $\edd=0$ we recover the established result for the non-dipolar system, $\omega_{\rm s} \approx \omega_z/\sqrt{2}$ \cite{busch_2000}.  More generally, $\omega_{\rm s}$ varies sensitively with $\edd$, deviating by up to $40\%$ from the non-dipolar frequency.  In comparison, for $\edd=0$, the deviation from $\omega_z/\sqrt{2}$ is only significant in the very weakly-interacting limit $\beta \lappeq 1$; for example, a non-dipolar system with comparable condensate and soliton sizes to Fig. \ref{fig:osceg05}(a) oscillates to within $5\%$ of $\omega_z/\sqrt{2}$.  The scale of this sensitivity is surprising given that the other collective oscillations - the shape oscillations - in elongated dipolar BECs vary much more weakly with $\edd$ (see, e.g. Fig. 11(a) of Ref. \cite{bijnen_2010}).

The intuitive explanation of this deviation is that the soliton, whose dynamics are governed by the curvature of the BEC profile, feels an effective trap frequency due to magnetostriction.  We can deduce this effective frequency by relating the dipolar TF profile, Eq. (\ref{eqn:diptf}), to an equivalent non-dipolar ($a_{\rm eff} \mapsto a_s$) TF profile with modified trap frequency ($\omega_z \mapsto \omega_{z, {\rm eff}}$).   This effective frequency is,
\begin{eqnarray}
\omega_{z,\text{eff}}=\frac{\omega_z}{\sqrt{1+\frac{1}{2}\edd[1-3\cos^2\theta]}}.
\label{eqn:mag}
\end{eqnarray}
Within this picture, the dark soliton would oscillate at $\omega_{\rm s}=\omega_{z,\text{eff}}/\sqrt{2}$.  However, this prediction (dotted line in Fig. \ref{fig:freqsig05}) fails to model the behaviour of $\omega_{\rm s}$.  This discrepancy is not accounted for by the effective mass of the soliton: the denominator of $\sqrt{2}$ in the predicted non-dipolar oscillation frequency is related to the soliton having an effective mass of $M_{\rm s}=2m$. We have evaluated, and corrected for, the effective mass of the soliton, as per Ref. \cite{edmonds_2016}, and find no significant effect.   The incapability of the particle model, so successful for non-dipolar dark solitons, to describe the observed oscillations leads us to conclude that the dipolar dark solitons are inherently extended and non-particle-like excitations, which cannot be decomposed from the background BEC.

To assess the role of dimensionality, we have conducted the corresponding simulations using the full 3D dipolar GPE \cite{code}.   The dimensional stability of the dark solitons in this system is confirmed.  Moreover, the 3D oscillation frequencies (red circles in Fig. \ref{fig:freqsig05}) are similar to the 1D results, although the RI regimes encroach to lower $\edd$ in 3D.  For example, the 3D BEC is stable for $-10 \lappeq \edd \lappeq -3$ for $a_s<0$ and $-2 \lappeq \edd \lappeq 16$ for $a_s>0$.  The decreased stability in 3D is due to the role of transverse magnetostriction in facilitating the RI \cite{Bohn2009}; indeed, as the ratio $\omega_z/\omega_\perp$ is decreased (system made more elongated), the RI is suppressed and approaches the 1D behaviour.

\begin{figure}[t]
	\centering
	\includegraphics[scale=0.9]{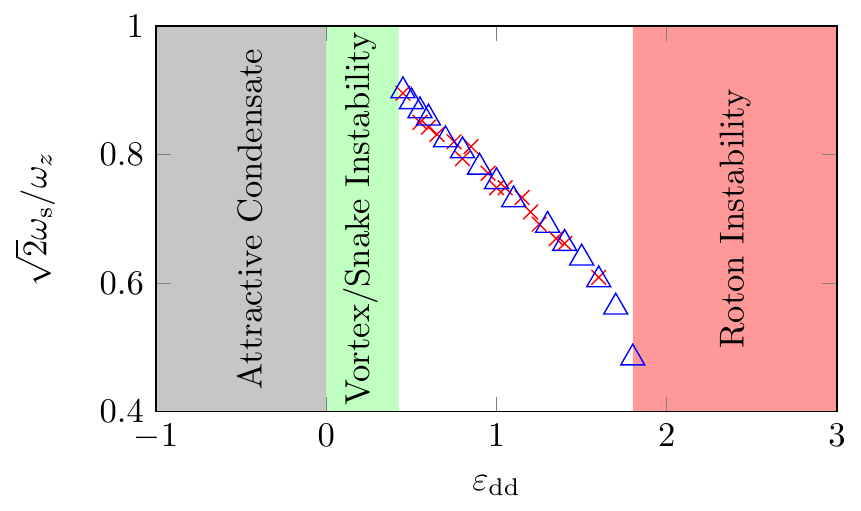}
	\caption{(Color online) Oscillation frequency and phase diagram for a dark soliton in a $^{164}$Dy BEC with Feshbach tuning of $a_s$, based on a recent experiment set-up \cite{ferrier_2016}.  Shown are cases where the soliton is imposed in the initial condition (blue triangles, as per Fig. \ref{fig:spec}) and imprinted in real time (red crosses).  Outside of the dark soliton regime, the condensate is either roton unstable (red), an attractive condensate incapable of supporting dark solitons (grey), or the dark solitons are dimensionally unstable.  Parameters:~$\theta=\pi/2$, $(\omega_\perp,\omega_z)=2 \pi \times (128,2)$ Hz, $a_\text{dd}=132a_0$ and $N=10000$.}
	\label{fig:exp_fig}
\end{figure}

Finally, we use 3D simulations to examine the dark solitons achievable in the elongated system of a recent experiment \cite{ferrier_2016} with $^{164}$Dy atoms ($a_{\rm dd}=132 a_0$), $\theta=\pi/2$ and $(\omega_\perp,\omega_z)=2 \pi \times (128,2)$ Hz, with variations of $\edd$ achieved through Feshbach tuning of $a_s$ (this is distinct to our previous results where $\beta$ was fixed).    Alongside introducing an off-centre black soliton into the initial condition (as done so far), we also imprint a $\pi$-phase step in real time, akin to experimental engineering of dark solitons \cite{Denschlag2000,Becker2008}.  This generates a soliton plus sound waves.  The oscillation frequency and phase diagram is depicted in Fig. \ref{fig:exp_fig}.  Stable dark solitons are supported for $0.4 \lappeq \edd \lappeq 1.8$; across this range $\omega_{\rm s}$ changes by a factor of two.  Above this range, the RI arises, matching the observed experimental value of condensate collapse for this system.  Below this range, the dark solitons undergo the snake instability into vortex rings.  This is because the regime of small positive $\edd$ corresponds to large positive $a_s$ and hence a small healing length; when this becomes smaller than the transverse system size, the condensate leaves the quasi-1D regime and becomes 3D in nature.   For negative $\edd$, i.e. negative $a_s$, the large and attractive contact interactions dominate to form a net attractive trapped condensate, in which dark solitons are not supported.

In conclusion, dark solitons are supported in trapped quasi-1D dipolar BECs, providing the background BEC is itself stable and net repulsively-interacting.  These excitations are accessible to current experiments.  While dark solitons in non-dipolar trapped BECs oscillate as classical particles at a characteristic and robust (e.g. insensitive to interactions) ratio of the trap frequency, dipolar interactions shatter this behaviour.  The oscillation frequency depends sensitively on the dipolar interactions, with the dependence being remarkably larger than predicted for the other collective excitations - the shape modes.  Similar anomalous oscillations have been predicted across the BEC-BCS crossover in Fermi gases \cite{scott_2011}.  Moreover, the particle model fails to capture the oscillation frequency, even accounting for magnetostriction of the BEC; the implication is that the dark soliton is strictly an extended and non-particle-like excitation.  These states offer a novel platform to study non-local dark solitons, to date observed in optics \cite{dreischuh_2006} and liquid crystals \cite{piccardi_2011}, with the immense control afforded by the atomic physics toolbox.  Finally, our results show that this species of quantum canary \cite{Anglin2008} is particularly sensitive to the interactions, suggest their potential use to probe the mesocopic details of the quantum field, such as the current open questions over quantum fluctuations in dipolar BECs \cite{chomaz_2016,schmitt_2016,bisset_2016,wachtler_2016}.

\noindent {\it Acknowledgements} - TB, ME and NGP thank the Engineering and Physical Sciences Research Council (Grant No. EP/M005127/1) for support. KP and KR thank the (Polish) National Science Center grant no. DEC-2012/04/A/ST2/00090 for support.

\end{document}